\documentclass[reprint,amsmath,amssymb,aps,pre,floatfix]{revtex4-2}

\usepackage{etoolbox}
\usepackage{array}
\usepackage{graphicx}
\usepackage{amsmath,amssymb,amsfonts}
\usepackage{amsthm}
\usepackage{booktabs}
\usepackage{algorithm}
\usepackage{algorithmicx}
\usepackage{algpseudocode}
\usepackage{listings}
\usepackage{dcolumn}
\usepackage{bm}
\usepackage[utf8]{inputenc}
\usepackage[T1]{fontenc}
\usepackage{relsize} % for mathlarger
\usepackage{subcaption} % for figure alignment
\usepackage{caption}
\usepackage{bookmark}
\providecommand{\bmhead}[1]{\section*{#1}}
\usepackage{adjustbox}
\usepackage{xcolor}

\usepackage{soul}
\usepackage{float}
\usepackage{hyperref}
\hypersetup{
    colorlinks=true,   % Enable colored links
    linkcolor=blue,    % Color for internal links
    citecolor=blue,     % Color for citations
    urlcolor=blue      % Color for URLs
}
\newtheorem{theorem}{Theorem}
\begin{document}
\preprint{APS/123-QED}

\title{IPSR Model: Misinformation Intervention through Prebunking in Social Networks}

\author{Robert Rai$^1$}
\author{Rajesh Sharma$^{2,3}$}
\author{Chandrakala Meena$^1$}
\email{chandrakala@iiserpune.ac.in}

\affiliation{$^1$Department of Physics, Indian Institute of Science Education and Research, Pune 411008, India}
\affiliation{$^2$Plaksha University, Sahibzada Ajit Singh Nagar, Punjab, 140306, India}
\affiliation{$^3$Institute of Computer Science, University of Tartu, Ülikooli 18, 50090, Tartu, Estonia}

\begin{abstract}

The rapid dissemination of misinformation through online social networks poses a growing threat to public understanding and societal stability. \textit{prebunking}, a proactive strategy based on inoculation theory, has recently emerged as an effective intervention to build cognitive resilience against misinformation before exposure. In this work, we investigate the impact of \textit{prebunking} on misinformation dynamics using a compartmental modeling framework. We first analyze the classical Ignorant–Spreader–Stifler (ISR) model, it's parameters are determined using empirical rumor data from Twitter. We then propose an extended model, the Ignorant–Prebunked–Spreader–Stifler (IPSR) model, which incorporates \textit{prebunking} as a preventive state and includes a forgetting mechanism to account for the decay of cognitive immunity over time. Using mean-field approximations, we derive steady-state solutions and examine the effect of \textit{prebunking} on the spreading of misinformation. We further investigate the robustness of the IPSR model by varying network size and average degree. In addition, we analyze the model's behavior on Watts–Strogatz and Barabási–Albert networks to assess the role of small-world and scale-free structures in shaping intervention outcomes. Our results show that the inclusion of \textit{prebunking} significantly reduces the scale of misinformation outbreaks across different network structures. These findings highlight the efficacy of \textit{prebunking} as a scalable intervention strategy and underscore the utility of compartmental models in understanding and mitigating information-based contagion in complex networks.

\end{abstract}

\maketitle

\section{Introduction}\label{sec1}
In the present era, digital information spreads faster than ever before. This rapid flow of information, while beneficial in many ways, also carries significant risks, particularly regarding misinformation. Misinformation refers to content that is false or misleading, whether it is shared intentionally or not. The impact of misinformation can be profound, undermining public understanding, jeopardizing safety, and influencing crucial decision-making processes. Social networks amplify this challenge, as their algorithms often prioritize sensational or divisive content, making misinformation highly visible and readily accessible\cite{centola2010spread,del2016spreading,lazer2018science,chen2023spread,duzen2023analyzing}. Digital misinformation on social media has become so widespread that the World Economic Forum (WEF) now considers it a significant threat of the $21^{st}$ century \cite{WEF2024}. Counter responses, such as debunking or fact-checking, which are reactive, lack the timeliness or scope needed to effectively counter misinformation once it has already reached a broad audience \cite{wang2018rumor,himma2022debunking}. It has been observed that after debunking, people's perceptions remain shaped by the continuing influence of misinformation, which makes debunking less effective \cite{johnson1994sources,seifert2002continued,lewandowsky2017beyond}. As a result, misinformation intervention studies have shifted toward preventive strategies, such as \textit{prebunking}. It is one of the most promising approaches to inoculate individuals against the influence of false information before encountering it \cite{roozenbeek2020prebunking}.

\subsection*{\textit{Prebunking}}
\textit{Prebunking}, rooted in inoculation theory\cite{compton2024inoculation}, applies a psychological principle similar to vaccination. This theory was introduced by William J. McGuire in the 1960s and says that exposing people to weakened forms of challenges could enhance their resilience to future attempts at persuasion \cite{mcguire1964inducing}.
This preventive approach helps individuals develop \textquotedblleft cognitive immunity\textquotedblright, equipping them with mental defenses against manipulation \cite{pfau1997enriching,lewandowsky2021countering}. In essence, \textit{prebunking} involves pre-exposing individuals to typical misinformation tactics or weakened versions of misleading narratives with counter-arguments to build psychological resilience. When these individuals later encounter actual misinformation, they are more likely to critically assess it, reducing the likelihood of its spread. This process has shown considerable success, particularly when \textit{prebunking} content is tailored to specific social contexts and reinforced over time \cite{banas2010meta, bavel2020using}. The role of analytic thinking in misinformation resistance demonstrates that individuals who engage in more reflective, analytical thinking are less likely to believe false information \cite{pennycook2019fighting}. This insight supports the goals of \textit{prebunking}, which seeks to encourage critical thinking.

Recent global events have underscored the need for effective \textit{prebunking} methods, which affect several processes. For instance, during the COVID-19 pandemic, misinformation related to health risks, treatments, and preventative measures associated with various vaccinations spread rapidly \cite{bonnet2020covid, caceres2022impact}. It caused public confusion and hampered efforts to control the virus.
\textit{prebunking} campaigns launched during this period highlighted the potential of inoculation strategies in digital information spaces, with interventions such as educational videos and interactive games significantly improving people's critical thinking skills regarding misinformation \cite{goviral,van2020inoculating,basol2021towards}. Moreover,  \textit{prebunking} has been used to inoculate the public against misinformation about climate change \cite{van2017inoculating} and also to address election-related misinformation, where narratives can influence voter perceptions and democratic processes \cite{Google2024}. Another notable \textit{prebunking} initiative was launched by Google and Jigsaw\cite{Google2024} to build Resilience to Online Manipulation Tactics in Germany, Countering Anti-Refugee Narratives in Central \& Eastern Europe. Such applications showcase prebunking’s adaptability and relevance in diverse contexts, from health to politics, as misinformation continuously evolves to exploit public vulnerabilities.

The study of misinformation diffusion and counter strategies has evolved significantly, drawing on concepts from fields such as epidemiology\cite{keeling2005networks,kauk2021understanding}, psychology \cite{van2022misinformation,blair2024interventions,butler2024nudge}, and network science\cite{pastor2015epidemic,meena2017threshold}. The epidemic model provides a framework for understanding how a “contagion” spreads across a population\cite{ beckley2013modeling,schlickeiser2024mathematics}. Most misinformation spreading models are based on the epidemic model, and this approach has proven effective for studying misinformation dynamics \cite{zhao2011rumor,zhao2012sihr}. The epidemic model is adapted by further incorporating additional characteristics to better capture the complexities of information spread \cite{chen2020rumor,el2024controlling, schlickeiser2021analytical}. A recent study introduced a weak-immune model \cite{turkyilmazoglu2022extended}, which accounts for partial immunity. This serves as a useful analogy for understanding how \textit{prebunking} fosters a temporary, weakened resilience to misinformation.

The success of epidemiological models inspires the present work to quantitatively model \textit{prebunking} effects on misinformation resistance within complex social networks. By simulating information dynamics through a compartmental model, the study conceptualizes individuals as belonging to distinct states — ignorant, prebunked, spreader, and stifler — based on their exposure to \textit{prebunking} and misinformation. Initially, we introduce the ISR model \cite{jiang2021reciprocal} to understand the spreading pattern without considering \textit{prebunking}. The parameters in the ISR model are obtained by fitting the spreader population to a real-world rumor dataset from X (formerly Twitter). Then, we introduce the IPSR model and perform dynamical analysis for the steady state.
To incorporate real-world scenarios, we include the forgetting mechanism of the \textit{prebunking} information and perform sensitivity analyses by varying the rate of \textit{prebunking}, which results in a higher population being prebunked when misinformation dissemination begins, and by varying the degree of \textit{prebunking} effectiveness. We further perform simulations on Watts-Strogatz (WS) and Albert-Barabási (BA) networks to verify the consistency of the spreading dynamics. We observe how the variance in certain parameters affects the system's population in different classes and helps identify optimal strategies for designing and implementing effective \textit{prebunking} interventions in digital spaces.

Ultimately, this research contributes to a growing body of work that emphasizes preventive misinformation strategies \cite{pennycook2020implied, zarocostas2020fight}. Understanding the \textit{prebunking} effects on misinformation dynamics helps policy-makers, social media platforms, and public health officials design evidence-based interventions that mitigate the impact of misinformation. In doing so, this study supports the development of resilient digital environments where individuals are equipped to critically engage with both information and misinformation, thereby strengthening societal resilience against the proliferation of harmful content.

The rest of the paper is structured as follows: In Section \ref{sec2}, we introduce the ISR model, which includes only misinformation dissemination and does not account for \textit{prebunking}. The parameter estimation from the Twitter dataset is also presented in this section. In Section \ref{sec3}, we formulate the IPSR model. In Section \ref{sec4}, we present the corresponding analytical theorems. Section \ref{sec5} and Section \ref{sec6} contains numerical results and sensitivity analyses for the mean field approximation and heterogeneous networks, and Section \ref{sec7} offers a discussion of the findings, concluding remarks, and future directions.

\section{ISR Model: Misinformation Spreading without \textit{prebunking}}\label{sec2}
As described in the introduction, the spread of information in social networks can be studied using the compartmental framework, and there are many models that describe information spread. Initially, we exclude \textit{prebunking} and consider one such model to understand spreading dynamics, that is, the ISR model (Ignorant–Spreader–Stifler) \cite{jiang2021reciprocal}. In this model, the population of a social network is divided into three compartments: Ignorants$(I)$, Spreaders$(S)$, and Stiflers$(R)$, each representing the fraction of individuals of the total population who play different roles in the misinformation diffusion process. These classes of population are described as:

\begin{itemize}
    \item Ignorant ($I$): Individuals who do not know about the misinformation. These people are vulnerable to misinformation and can become a spreader or stifler upon contact with a spreader.
\begin{figure}[h]
    \includegraphics[width=0.8\linewidth]{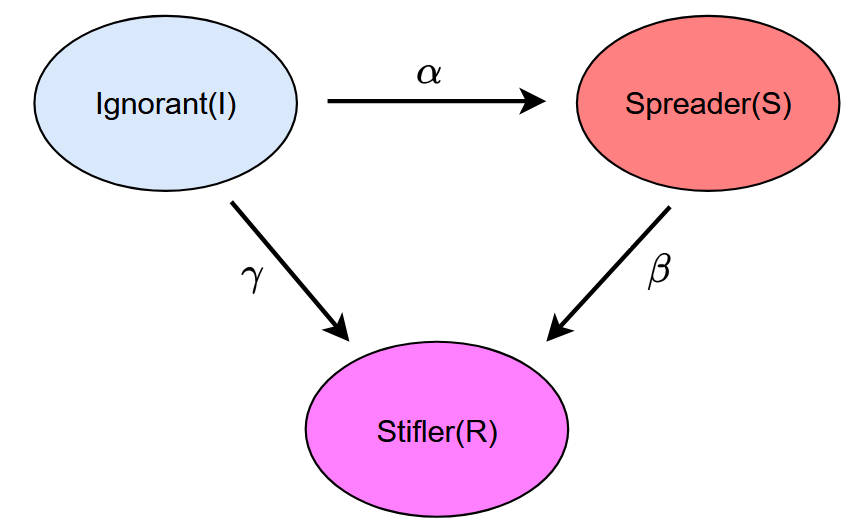}
    \caption{\textbf{ISR Model:} Transition diagram between Ignorants, Spreaders and Stiflers.}
    \label{Fig:isr}
\end{figure}
\begin{figure*}[htbp]
    \centering
   \includegraphics[width=0.9\textwidth]{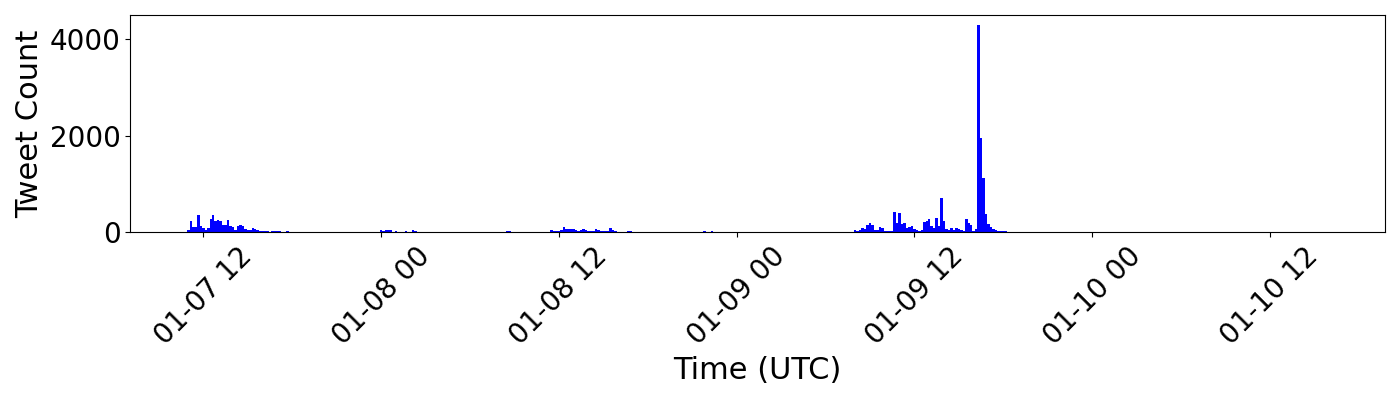}
    \caption{The tweet counts of the Charlie Hebdo event of the PHEME dataset from X (formerly Twitter) in bins of 10-minute intervals. The initial tweets started on 7th January 2015 and include tweets up to 10th January 2015.}
    \label{Fig:full_tweet_counts}
\end{figure*}
    \item Spreader ($S$): Individuals who know and actively spread the misinformation.
    \item Stifler ($R$): Individuals who learned about the misinformation but do not spread it either because they know the correct information or because they spread it a long time ago and later lost interest in continuing to do so.
\end{itemize}

The transitions among these compartments groups are illustrated in Fig.~\ref{Fig:isr}. At the initial stage $(t=0)$, when there is no significant misinformation activity, the majority of the population consists of Ignorants, with only one or a few individuals acting as Spreaders. When an Ignorant comes into contact with a Spreader, they may become a Spreader themselves at a transmission rate $\alpha$. Over time, Spreaders lose their interest or influence and transition into Stiflers at a rate $\beta$. Additionally, upon contact with a Spreader, an Ignorant may also become a Stifler - rather than a Spreader - at a rate $\gamma$, possibly because they disbelieve the information or deem it irrelevant.

For homogeneous networks, the mean-field equations that describe the dynamics of the population among the three compartments can be described as :
\begin{subequations}
    \begin{align}
    \frac{dI}{dt} &= -\alpha \langle k \rangle I S - \gamma \langle k \rangle I S \\
    \frac{dS}{dt} &= \alpha \langle k \rangle I S - \beta S \\
    \frac{dR}{dt} &= \gamma \langle k \rangle I S +  \beta S
    \end{align}
    \label{rumor}
\end{subequations}
where $\langle k \rangle$ represents the average degree of the network. The total population satisfies the normalization condition $N = I + S + R = 1$.

\subsection{Basic Reproduction Number for ISR model}\label{subsec31}

The basic reproduction number, $\mathcal{R}_0$, is the average number of secondary spreaders arising from a single spreader in a community of entirely susceptible populations. $\mathcal{R}_0$ is pivotal in evaluating the severity of an outbreak and the efficacy of various interventions. When $\mathcal{R}_0<1$, each spreader influences fewer than one other person on average, preventing the number of spreaders from increasing and leading to the eventual decline of misinformation spread. Conversely, if $\mathcal{R}_0>1$,  the misinformation is more likely to disseminate rapidly within the population, potentially leading to an epidemic of false information.

For the given ISR model Eq.\ref{rumor} that does not include \textit{prebunking}, a misinformation outbreak occurs if the number of spreaders increases,
\begin{align*}
    \frac{dS}{dt} &> 0\\
    \frac{\alpha \langle k \rangle I }{\beta}  &> 1
\end{align*}
At the onset of the misinformation outbreak, there is one initial spreader, while the rest of the population is Ignorant.
Considering any arbitrary initial fractions of the ignorant population $(I_0)$, we have the expression for the basic reproduction number.
\begin{equation}
     \mathcal{R}_0 = \frac{\alpha \langle k \rangle I_0}{\beta}.
     \label{repro_old}
\end{equation}

\subsection{Parameter estimation using PHEME dataset}\label{sec2.2}
We utilize the PHEME dataset\cite{zubiaga2016pheme} from X (formerly Twitter) to investigate the spreading dynamics and to estimate the parameters $(\alpha,\beta,\gamma)$ of the ISR model. The dataset includes source and reply tweets associated with nine distinct real-world events. It also contains the "who-follows-whom" file to define the network size and connectivity. For our analysis, we focus on the Charlie Hebdo event, as it comprises one of the largest subsets of tweets within the dataset.
To quantify the temporal evolution of the spread of information, we calculate the total number of tweets related to the event in 10 minutes intervals. The resulting time series is presented in Fig.~\ref{Fig:full_tweet_counts}.

\begin{figure}[H]
    \centering
    \includegraphics[width=\linewidth]{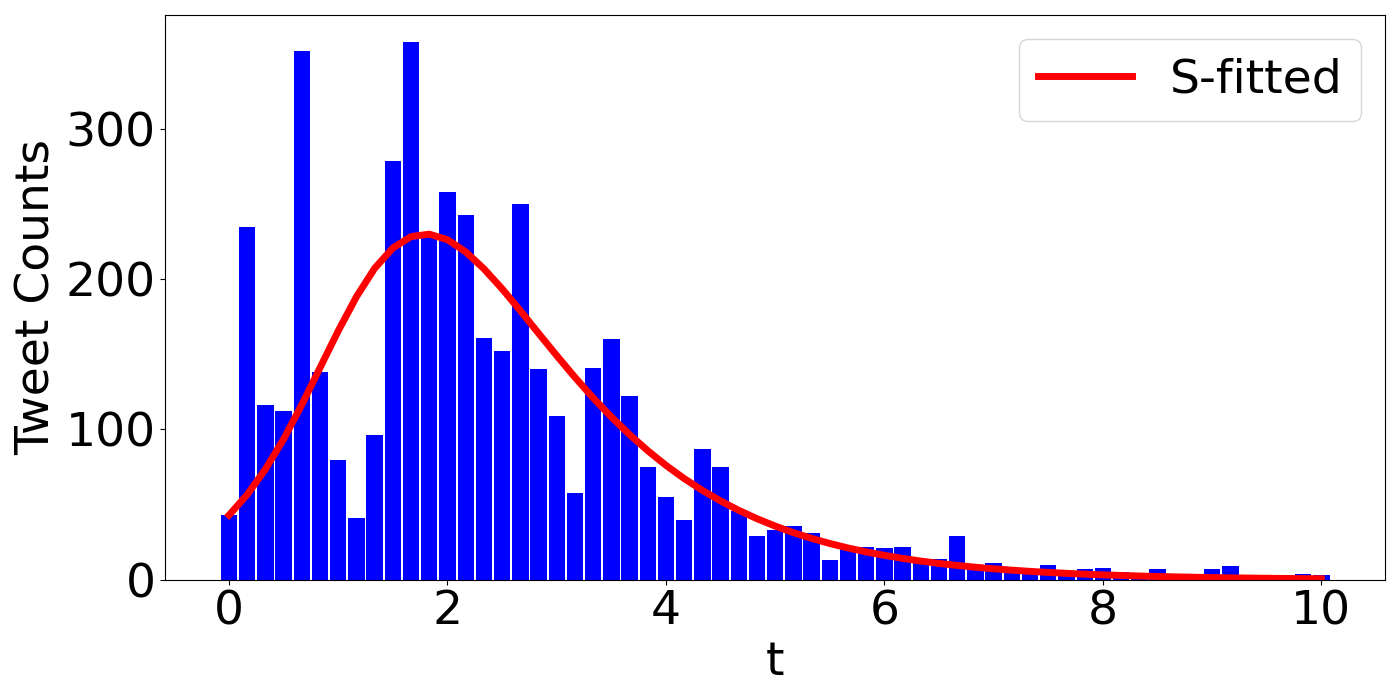}
    \caption{ Tweet counts for the first 10 hours for the Charlie Hebdo event and fitted ISR model for the $S$ class. All parameters are kept bound within $[0,1]$ to fit $S$ with the tweet counts.
    }
    \label{Fig:fitted_tweet_counts}
\end{figure}

The rumor tweets consist of multiple waves. We chose the first wave that lasted for 10 hours, which is considered to be a complete rumor spreading process. We estimated $\langle k \rangle = 3$ and $N = 3909$ for the Charlie Hebdo event from the given dataset, which is a small subset of the Twitter network. Every tweet that corresponds to the specific event counts as a Spreader, which is fitted to the ISR model for the Spreader population, shown in Fig.\ref{Fig:fitted_tweet_counts}. Keeping the parameter values bound between $0$ and $1$, we use the Python Scipy package to fit $S$ with the number of tweet counts, with minimum marginal error. We obtain $\alpha = 0.09$, $\beta = 0.1$, and $\gamma = 0.4$, which will be used for numerical analysis of the IPSR model in the later section.

\section{IPSR Model: Misinformation Spreading with \textit{prebunking}}\label{sec3}
In the IPSR model, the population of a network is categorized into two distinct groups before the spread of misinformation (Ignorant and Prebunked). When misinformation originates from a single source and spreads across a social network, two new compartments emerge, and the total population is divided into four different compartments. So, we develop a mathematical model consisting of the following four compartments: 

\begin{itemize}
    \item Ignorant (unaware) ($I$): Individuals who do not know about the \textit{prebunking} awareness and the misinformation. These people are at risk of either becoming spreaders of misinformation or being inoculated against it through \textit{prebunking}.
    \item Prebunked ($P$): Individuals who have received the \textit{prebunking} information and thus possess a weak psychological immunization. They are prone to revert back to the ignorant state when their \textit{prebunking} awareness wears off over time.
    \item Spreader ($S$) and Stifler ($R$): Both of these have the same description as in the ISR model.
\end{itemize}

\begin{figure*}[htbp]
    \centering
    \captionsetup{justification=justified, singlelinecheck=off}
    \includegraphics[width=0.7\linewidth]{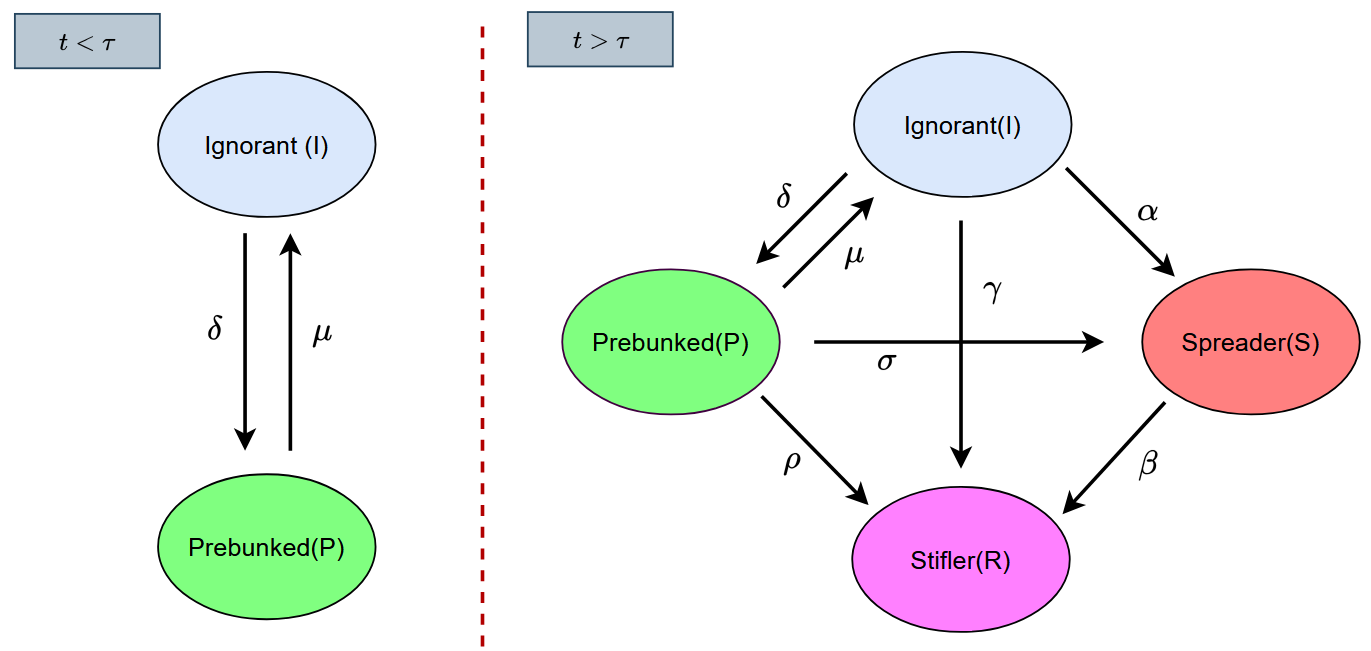}
    \caption{\textbf{Flow diagram for the IPSR model system.} When $t<\tau$, before the spreading of misinformation, the movement of individuals occurs only between $I$ and $P$ classes. We assume that at $t>\tau$, the spread of misinformation begins, leading to the introduction of two additional compartments: $S$ and $R$ representing spreader and stifler individuals, respectively.} 
    \label{Fig:model}
\end{figure*}

We consider that the total population $N$ is constant, $N=1$; hence, individual populations in each compartment will lie between 0 and 1, representing the fractions of the total populations in each compartment.
Fig.\ref{Fig:model} represents the flow diagram of population transition across the different compartments of the IPSR model. We assume that \textit{prebunking} is initiated before the spread of misinformation related to significant events such as elections or pandemics, etc\cite{Google2024} through ads, videos, or games on social platforms. Let $\tau$ be the time at which misinformation starts spreading.
The process that takes place before spreading misinformation ($t<\tau$) is shown left side of Fig. \ref{Fig:model} and modeled using Eq. \ref{eq:before} where I population transit to P compartment at a constant rate of $\delta$ due to \textit{prebunking}, and P population may revert back to the I compartment at a rate $\mu$ because we assume that individuals tend to forget the awareness information after a certain period \cite{basol2021towards}. 

When $t<\tau$, only \textit{prebunking} exists and there is no spreading of misinformation, and our model considers the dynamical equations:
\begin{subequations}
\begin{align}
    \frac{dI}{dt} &= -\delta I + \mu P\\
    \frac{dP}{dt} &= \delta I -\mu P
\end{align}
\label{eq:before}
\end{subequations}
with zero individuals in the $S$ and $R$ compartments implying to their non-existence. Until the introduction of misinformation, the population gets divided between ignorant and prebunked compartments such that the condition $I + P = 1$ holds true.

 At time  $t=\tau$, the dissemination of misinformation begins through a single spreader, and this leads to the introduction of two more compartments, namely Spreader ($S$) and Stifler $R$. Individuals move across compartments at certain rates that are specific between the two compartments (Table \ref{table1}).  
The above-mentioned process is modelled using eq. \ref{md:model} and is shown using the flow diagram in Fig.\ref{Fig:model} (right). 
The transition rules of individuals among the different compartments are as follows:
\begin{itemize}
    \item Ignorant individuals are given cognitive inoculation at the constant rate $\delta$. Meanwhile, prebunked individuals forget about the \textit{prebunking} information and revert back to ignorant at the rate $\mu$. This process remains unchanged even when misinformation is introduced into the network.
    
    \item When an ignorant individual comes into contact with a spreader, they can either become a spreader themselves with the rate $\alpha$, propagating the misinformation or choose not to spread it, becoming a stifler with the rate $\gamma$.
    
    \item When a prebunked individual comes into contact with a spreader, they can become a spreader with the rate $\sigma$ if they choose to propagate the misinformation or become a stifler with probability $\rho$ if they decide not to spread it.
    
    \item A spreader will become a stifler at a constant rate of $\beta$ due to a loss of interest, perceiving the information as irrelevant, or assuming that everyone else has already heard the misinformation.
    
\end{itemize}

The transition from a prebunked class to misinformation spreaders can occur despite efforts to teach critical thinking and resilience, as some individuals may believe and unknowingly share false rumors.
The dynamics governing the spread of misinformation and \textit{prebunking} intervention can be expressed through a series of differential equations. Considering the diffusion of misinformation for $t>\tau$, the mean-field equations that describe the dynamics of the population in degree-homogeneous networks can be formulated as:

\begin{subequations}
    \begin{align}
    \frac{dI}{dt} &= -\delta I -(\alpha + \gamma) \langle k \rangle I S + \mu P \label{2a}\\
    \frac{dP}{dt} &= \delta I -(\sigma + \rho) \langle k \rangle P S - \mu P \label{2b}\\
    \frac{dS}{dt} &= \alpha \langle k \rangle I S + \sigma \langle k \rangle P S - \beta S \label{2c}\\
    \frac{dR}{dt} &= \rho \langle k \rangle P S + \gamma \langle k \rangle I S +  \beta S \label{2d}
    \end{align}
    \label{md:model}
\end{subequations} 

The fractions of the population satisfy the normalization condition:
$ I+P+S+R = 1.$
We assume that the spread of misinformation originates from a single source at the outset. 
If the total population is $N$, the misinformation diffusion has the initial conditions:
\begin{align*}
    &I(0) = I_0, \hspace{4em} P(0) = P_0=1-I_0 - \frac{1}{N}\\
    &S(0) = \frac{1}{N}, \hspace{3.5em} R(0) = 0.
\end{align*}
where $I_0$ and $P_0$ are the initial populations in the ignorant and prebunked compartments when misinformation spreads.

\begin{table}[h]
\centering
\renewcommand{\arraystretch}{1.4}
\small
\begin{tabular}{|>{\centering\arraybackslash}p{5em}|p{20em}|}
 \hline
 \textbf{Parameter} & \multicolumn{1}{c|}{\textbf{Definition}} \\
 \hline
 $\alpha$  & the rate at which an ignorant becomes a spreader \\
 \hline
 $\beta$  & the rate that a spreader becomes a stifler over time \\
 \hline
 $\gamma$  & the probability that an ignorant individual directly turns into a stifler upon interaction with a spreader \\
 \hline
 $\delta$  & the \textit{prebunking} rate, shifting ignorant to prebunked class \\
 \hline
 $\mu$  & forgetting rate of \textit{prebunking} information \\
 \hline
 $\sigma$  & the rate at which prebunked becomes spreader \\
 \hline
 $\rho$  & the probability that a prebunked individual directly turns into a stifler upon contact with a spreader \\
 \hline
\end{tabular}
\caption{Description of the IPSR Model parameters}
\label{table1}
\end{table} 

\section{Dynamical Analysis of the IPSR model}\label{sec4}
Investigating the stability of the misinformation propagation model is crucial for developing effective control measures. This section provides a thorough analysis of the dynamic properties of the proposed IPSR model, focusing on the equilibrium points and their global stability. First, we calculate the basic reproduction number $\mathcal{R}_0'$, determining the conditions for spreading to die out or persist. Then, we confirm the positivity to validate our model, and finally, we analyze the stability conditions of the model.

\subsection{Reproduction Number for the IPSR Model}
In the IPSR model, the prebunked population remains susceptible to misinformation, albeit to a lesser extent. For a misinformation outbreak to occur, the expression should hold true,
\begin{align*}
    \frac{dS}{dt} > 0\\
    \alpha \langle k \rangle I S + \sigma \langle k \rangle P S &- \beta S > 0\\
    \frac{\alpha \langle k \rangle I  + \sigma \langle k \rangle P }{\beta}  &> 1
\end{align*}

Assuming any arbitrary initial fractions of the ignorant population ($I_0$) and the prebunked population ($P_0$) when the misinformation dissemination starts, we have the following inequality
\begin{equation*}
    \frac{\alpha \langle k \rangle I_0 + \sigma \langle k \rangle P_0}{\beta} = \mathcal{R}_0' > 1.
\end{equation*}
$\mathcal{R}_0'$ exhibit the same form as that of the weak-vaccination model\cite{turkyilmazoglu2022extended}.

The basic reproduction number can also be represented as :
\begin{equation}
    \mathcal{R}_0' = \frac{\alpha\langle k \rangle}{\beta + (\alpha-\sigma)\langle k \rangle P_0}
    \label{repro_formula}
\end{equation}
where we approximate $I_0 \approx 1- P_0$ for large $N$.

As the value of $\sigma$ approaches $\alpha$ or in the absence of \textit{prebunking}$(P_0 = 0)$, the second term in the denominator (Eq. \ref{repro_formula}) vanishes and it aligns with the reproduction number of ISR model (Eq. \ref{repro_old}) with $I_0 \approx 1$. 

\begin{figure}[h]
    \centering
    \includegraphics[width=0.9\linewidth]{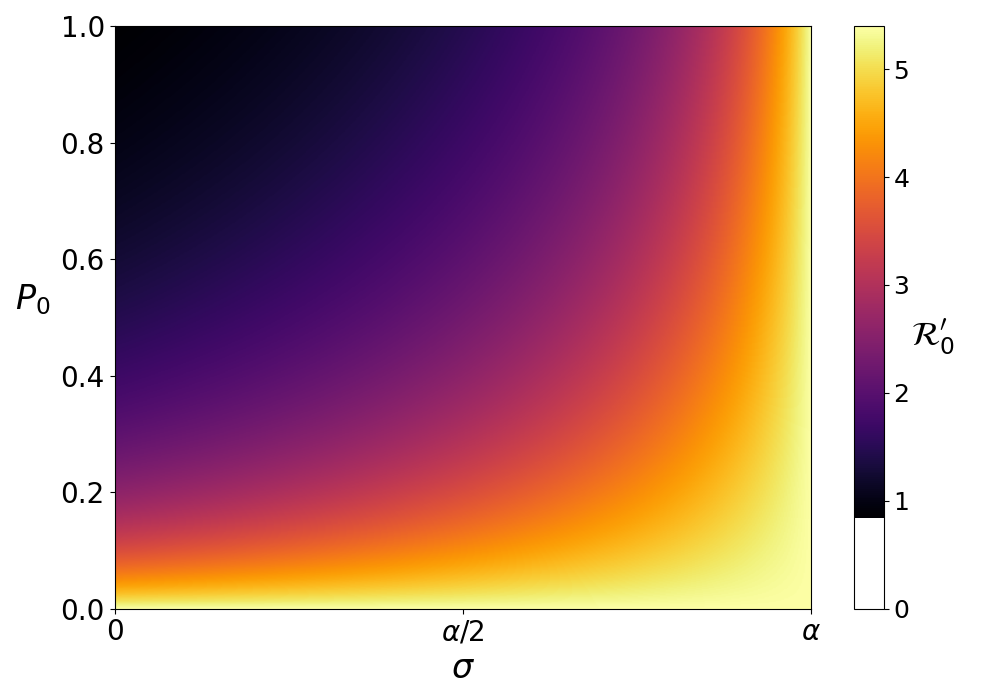}
    \caption{\textbf{Basic reproduction number $\mathcal{R}_0'$}. Effect of the initial prebunked population $P_0$ and $\sigma$ on the $\mathcal{R}_0'$. For $\mathcal{R}_0' <1$, there is a decline in the spread of misinformation, and it will eventually die out. Conversely, $\mathcal{R}_0' >1 $ suggests an epidemic of misinformation. The prebunked class has a lower probability of becoming a spreader, and thus the values of $\sigma$ are bound within $[0,\alpha]$.}
    \label{Fig:repro}
\end{figure} 

Fig.\ref{Fig:repro} represent a 2D heat map plotted by varying fractions of population, $P_0$ and $\sigma \in [0,\alpha]$ with $\langle k \rangle = 10$. When $P_0 \to 1$ and $\sigma \to 0$ (no transition from prebunked to spreader), the basic reproduction number $\mathcal{R}_0'$ becomes less than 1 highlighting that effective \textit{prebunking} to the vast majority of the population results in no emergence of a significant number of new spreaders.

\begin{figure*}[!htbp]
    \centering
    \begin{subfigure}[b]{0.32\textwidth}
        \centering
        \includegraphics[width=\linewidth]{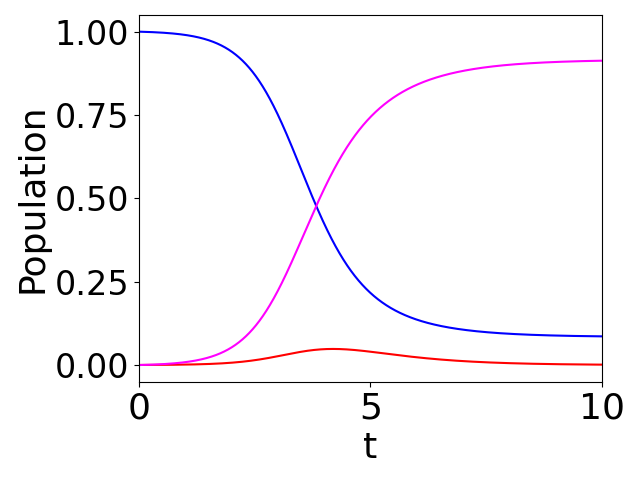}
        \caption{}
        \label{Fig:full_simulate(a)}
    \end{subfigure}
    \begin{subfigure}[b]{0.67\textwidth}
        \centering
        \includegraphics[width=\linewidth]{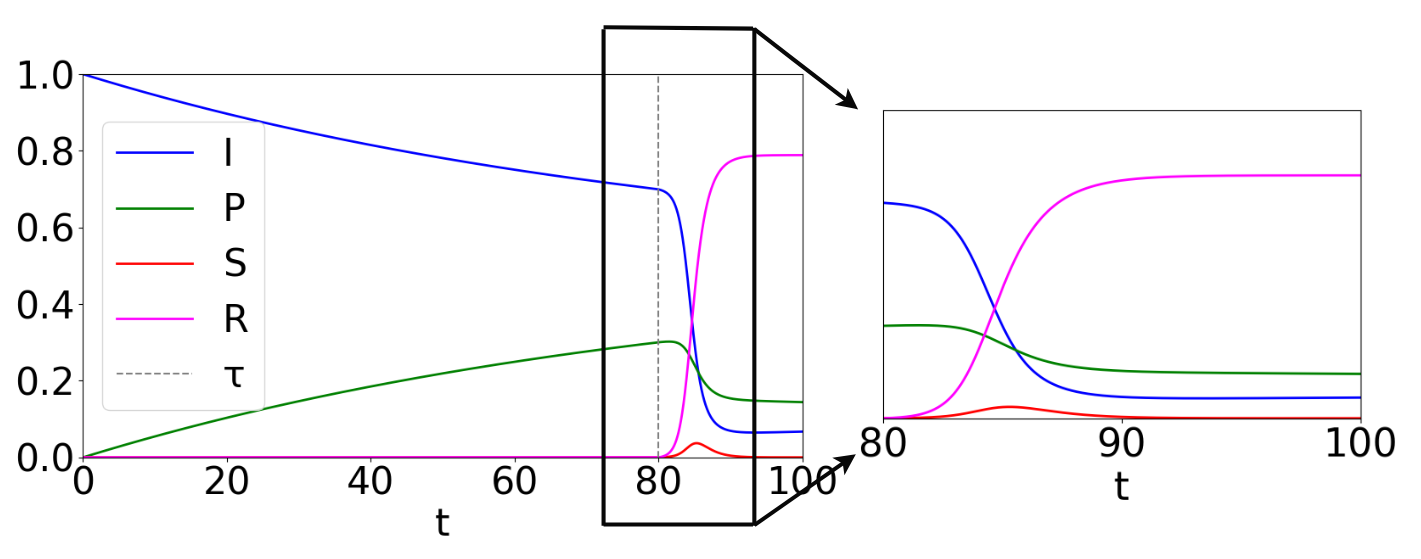}
        \caption{}
        \label{Fig:full_simulate(b)}
    \end{subfigure}

    \caption{\textbf{Temporal evolution of different classes of population in (a) ISR and (b) IPSR model}. The model parameters are $\alpha  = 0.09, \beta = 0.1, \gamma = 0.4, \delta = 0.00058, \mu = 0.0006, \sigma = 0.5\alpha, \rho = \beta$. Here $\tau = 80$ to obtain 30\% prebunked population when misinformation dissemination starts. 
    }
    \label{Fig:3}
\end{figure*}

\subsection{Positive Solution of the model}\label{subsec32}
The IPSR model helps to understand the dynamic behaviour of different populations over time. To ensure with real scenario, it is essential that all system variables remain non-negative. This characteristic is vital for confirming the model's validity and reliability in practical situations.
\begin{theorem}
    Let $\Psi = (S,P,I,R) \in \mathbb{R}^4 : S(0)>0$, $P(0)>0$, $I(0)>0$, $R(0)>0$, then the solution $S(t)$, $P(t)$, $I(t)$, $R(t)$ of the model is positive for all $t>0$.
\end{theorem}

The proof of the theorem and the subsequent theorems from the next subsection are given in the appendices.

\subsection{Dynamical Stability of Steady State Analysis}\label{subsec33}
After a considerable period, the number of spreaders diminishes to zero, and the system dynamics reach a steady state. The steady state can be analyzed through stability analysis around the equilibrium point.

\begin{theorem}
    The system attains an equilibrium state under the conditions \\
    (a) $S=0$ and \\
    (b) $\delta I=\mu P$ with $I+P+R=1$ (where $0\leq I,P,R\leq 1$),\\ applicable for all positive, non-zero parameters. 
\end{theorem}

\begin{theorem}
The system achieves a stable equilibrium state when the given condition, along with the conditions of theorem 2, is satisfied.
     \begin{equation}
         P \leq \frac{\beta\delta}{(\alpha\mu+\sigma\delta)\langle k \rangle}
         \label{C2}
     \end{equation}
\end{theorem}
\begin{theorem}
    The IPSR model is globally asymptotically stable.
\end{theorem}

\section{Numerical Simulations}\label{sec5}

Numerical simulations were performed using the Runga-Kutta method to solve the set of ordinary differential equations for both models. We consider $N$ population and average degree $\langle k \rangle$ taken from the dataset analyzed in Section \ref{sec2.2}. For a single initial spreader, Fig.\ref{Fig:full_simulate(a)} illustrates the temporal evolution of different classes of the ISR model. This figure illustrates that as the system evolves, the population of ignorant individuals decreases while the number of stiflers increases. The spreaders initially rise to a peak value, then gradually decrease, eventually approaching an equilibrium state when the number of spreaders reduces to zero. In online social networks, the number of spreaders tends to be relatively small compared to the overall population. Hence the peak value of spreaders is a small fraction. As the number of spreaders diminishes to zero, the ignorant and the stifler population also reach an equilibrium state.

For the IPSR model, when $t<\tau$, ignorant 
Individuals are given constant \textit{prebunking}, and misinformation does not exist during this period. Eq. \ref{eq:before} defines this process. However, we are interested in the dynamics after misinformation is introduced in the system. In the period following the onset of misinformation dissemination, the magnified region in Fig.\ref{Fig:full_simulate(b)} depicts the temporal dynamics of the IPSR model for the same population and the corresponding parameter settings. 
As per our model, the diffusion of misinformation starts at $t=\tau$ denoted in all the subsequent plots by dash vertical line. To compare population dynamics with the ISR model, we set $\tau = 80$ and obtain $30\%$ of the population in the prebunked compartment. We assume that the Ignorant and Prebunked classes have the same stifling rate, $\beta = \rho$.
Also the Prebunked class is assumed to have $50\%$ less chance of becoming a spreader, $\sigma = 0.5 \alpha$. The \textit{prebunking} Process and Forgetting mechanism are comparatively much slower that usually take days, and thus we set their rates $\delta$ and $\mu$ to a minimal value. As illustrated in Fig.\ref{Fig:full_simulate(b)}, there is a reduction in the peak value of spreaders (by $23\%$) and stifler population when \textit{prebunking} is considered. This decrease can be attributed to the fact that fewer individuals transition to the Spreader class from the Prebunked class, resulting from cognitive immunization diminishing the likelihood of influence from misinformation.

\begin{figure}[h]
    \centering
    \captionsetup{justification=justified, singlelinecheck=off}
    \includegraphics[width=0.9\linewidth]{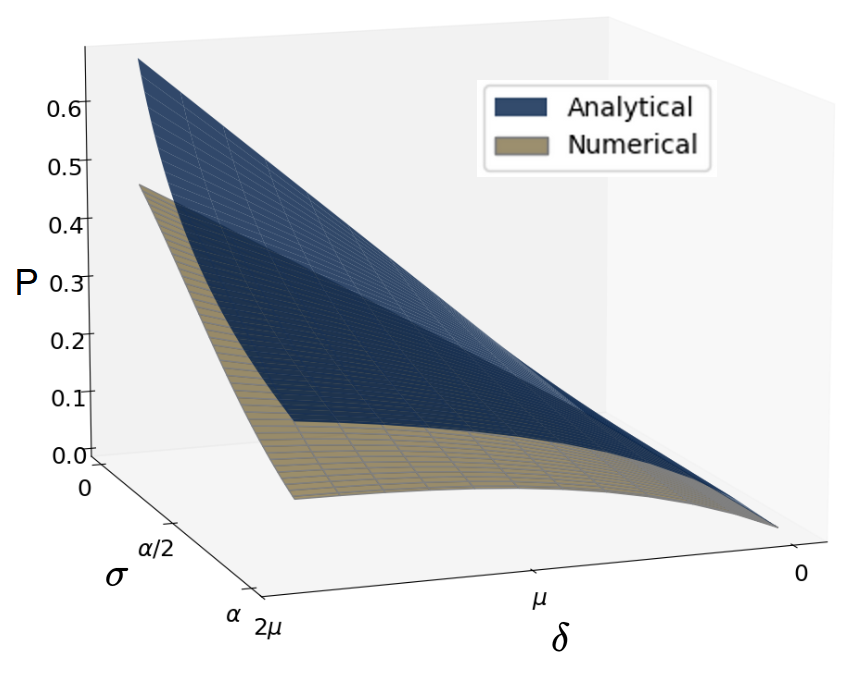}
    \caption{Analytical and Numerical Simulation: The simulated values of P remain less than the proposed analytically upper bound for the subset of the model parameters.}
    \label{proof}
\end{figure}

\textbf{\textit{Numerical Verification}}: In Theorem \textbf{3}, we proposed an upper bound for the steady-state value of 
$P$, which depends on a subset of the model parameters. Specifically, the maximum attainable value of $P$ is,
\begin{equation*}
    \displaystyle P \leq \frac{\beta\delta}{\big(\alpha\mu + \sigma\delta\big)\big\langle k \big\rangle}
\end{equation*}
Keeping the other parameters consistent, we plot the maximum analytical value of $P$ by varying $\sigma$ and $\delta$. The steady state of $P$ obtained from simulation for the same set of parameters is also plotted, and it is shown in Fig.\ref{proof}.
The simulated values of $P$ for any set of parameters remain less than the maximum value and thus verify the analytical proposition.
\begin{figure*}
    \centering
    \begin{subfigure}[b]{0.24\textwidth}
        \centering
        \includegraphics[width=\linewidth]{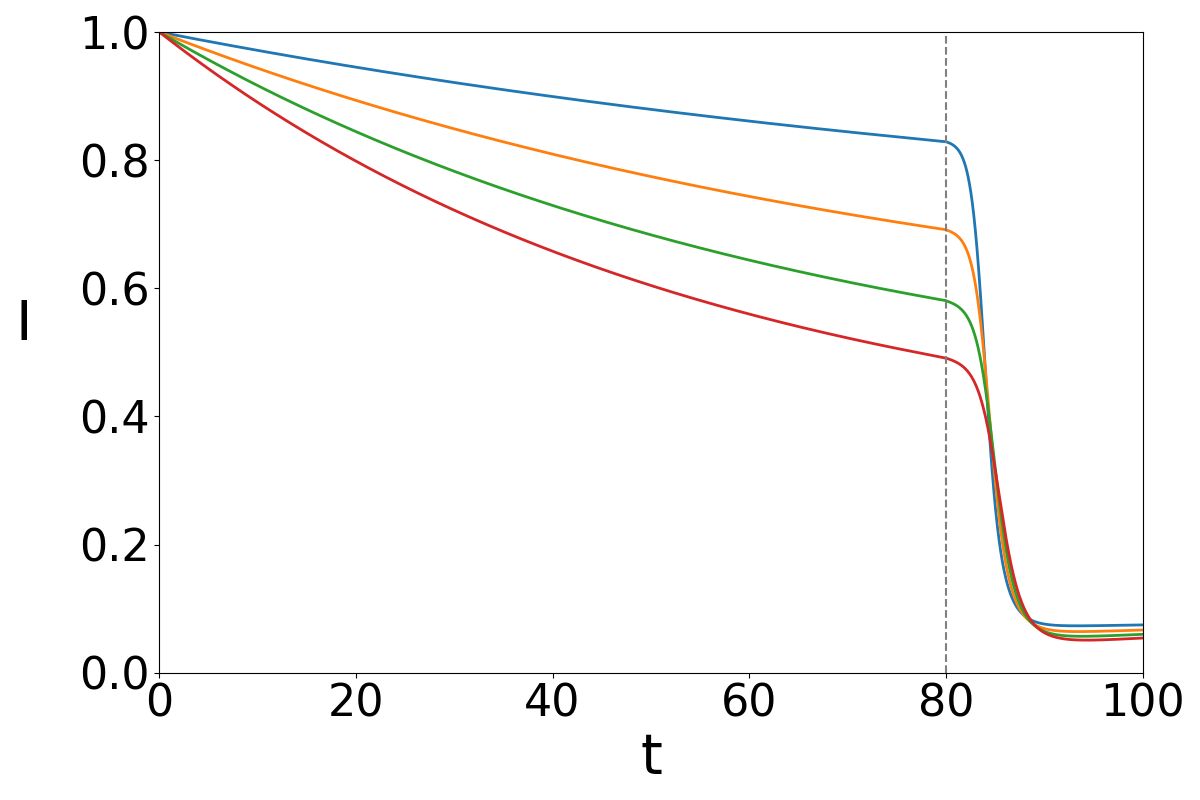}
        \caption{}
        \label{Fig:rate(a)}
    \end{subfigure}
    \hfill
    \begin{subfigure}[b]{0.24\textwidth}
        \centering
        \includegraphics[width=\linewidth]{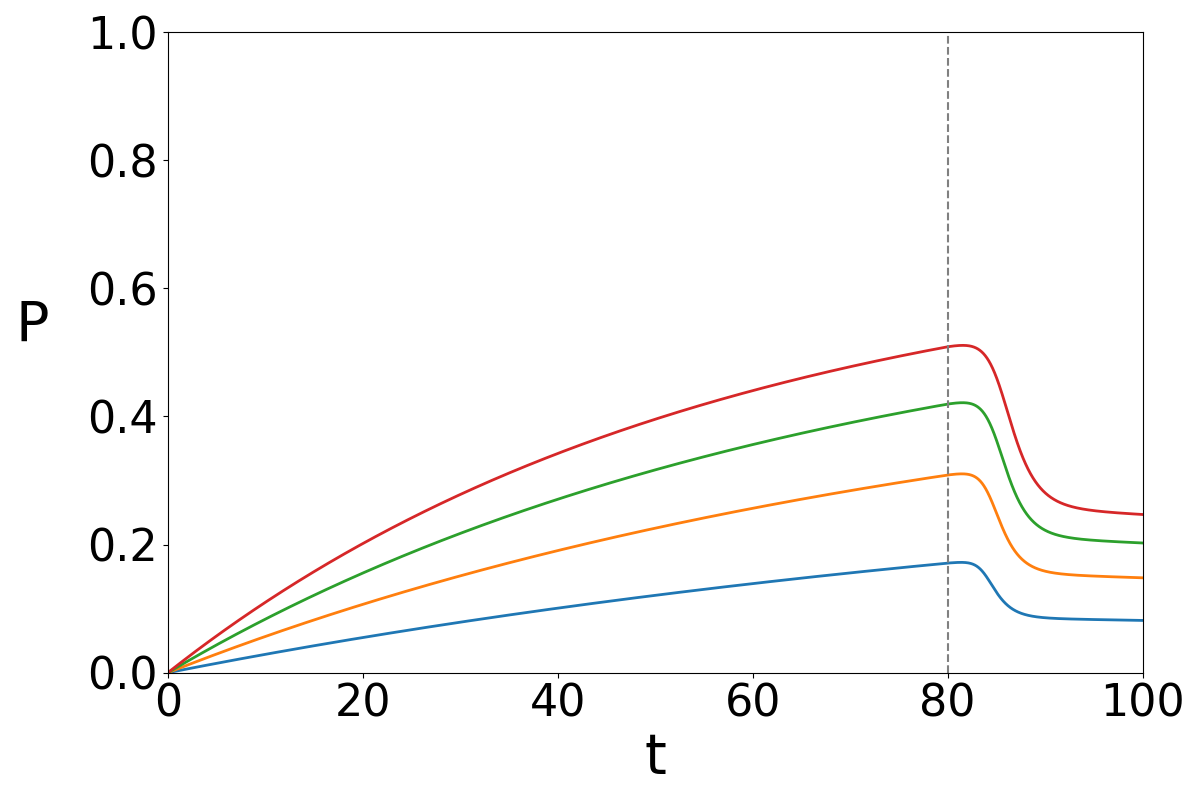}
        \caption{}
        \label{Fig:rate(b)}
    \end{subfigure}
    \hfill
    \begin{subfigure}[b]{0.25\textwidth}
        \centering
        \includegraphics[width=\linewidth]{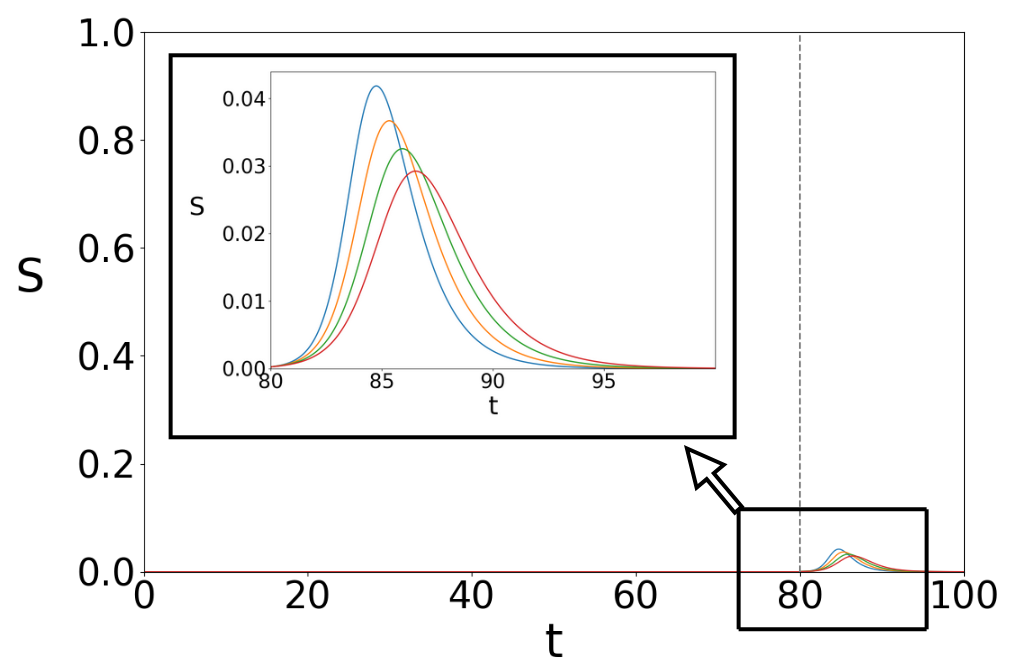}
        \caption{}
        \label{Fig:rate(c)}
    \end{subfigure}
    \hfill
    \begin{subfigure}[b]{0.24\textwidth}
        \centering
        \includegraphics[width=\linewidth]{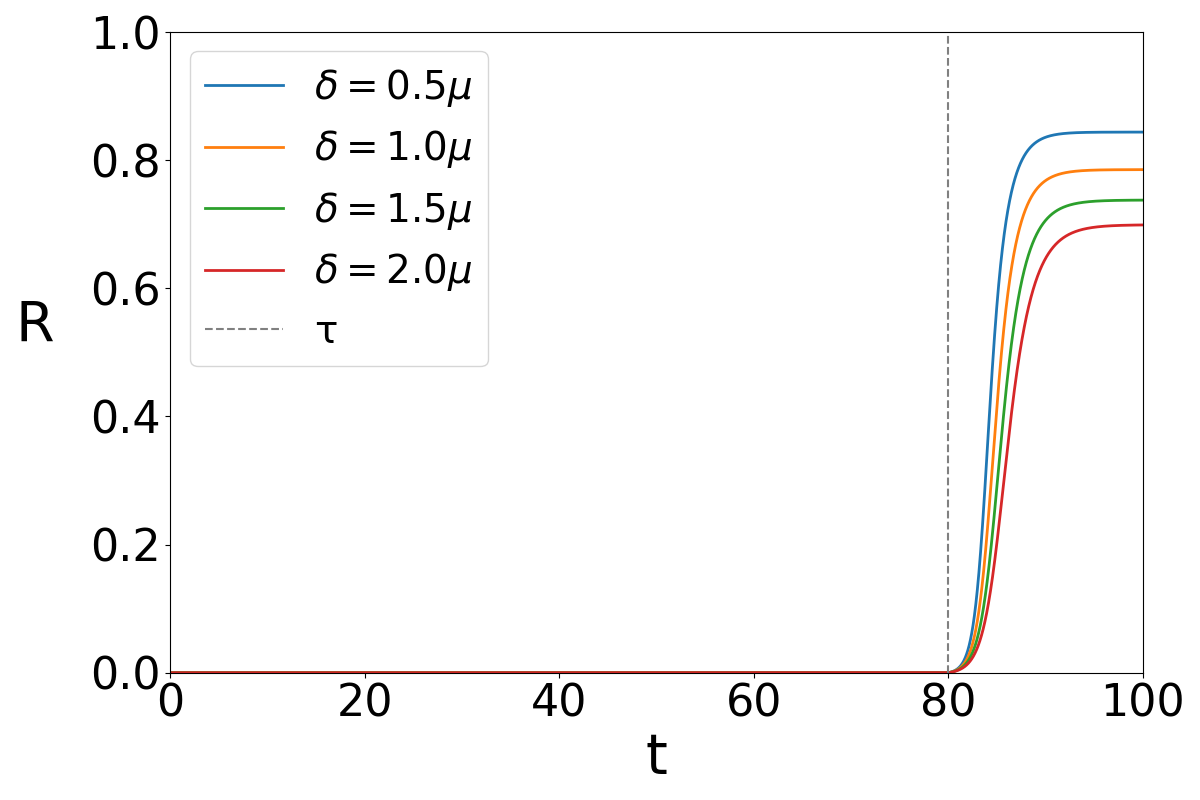}
        \caption{}
        \label{Fig:rate(d)}
    \end{subfigure}
    \caption{\textbf{Effect of \textit{prebunking} rate ($\delta$).} Influence of different $\delta$ on the temporal evolution of \hspace{4in} (a) $I$, (b) $P$, (c) $S$ and (d) $R$.
    }
    \label{Fig:rate}
\end{figure*}
\begin{figure*}[htbp]
    \centering
    \begin{subfigure}[b]{0.24\textwidth}
        \centering
        \includegraphics[width=\linewidth]{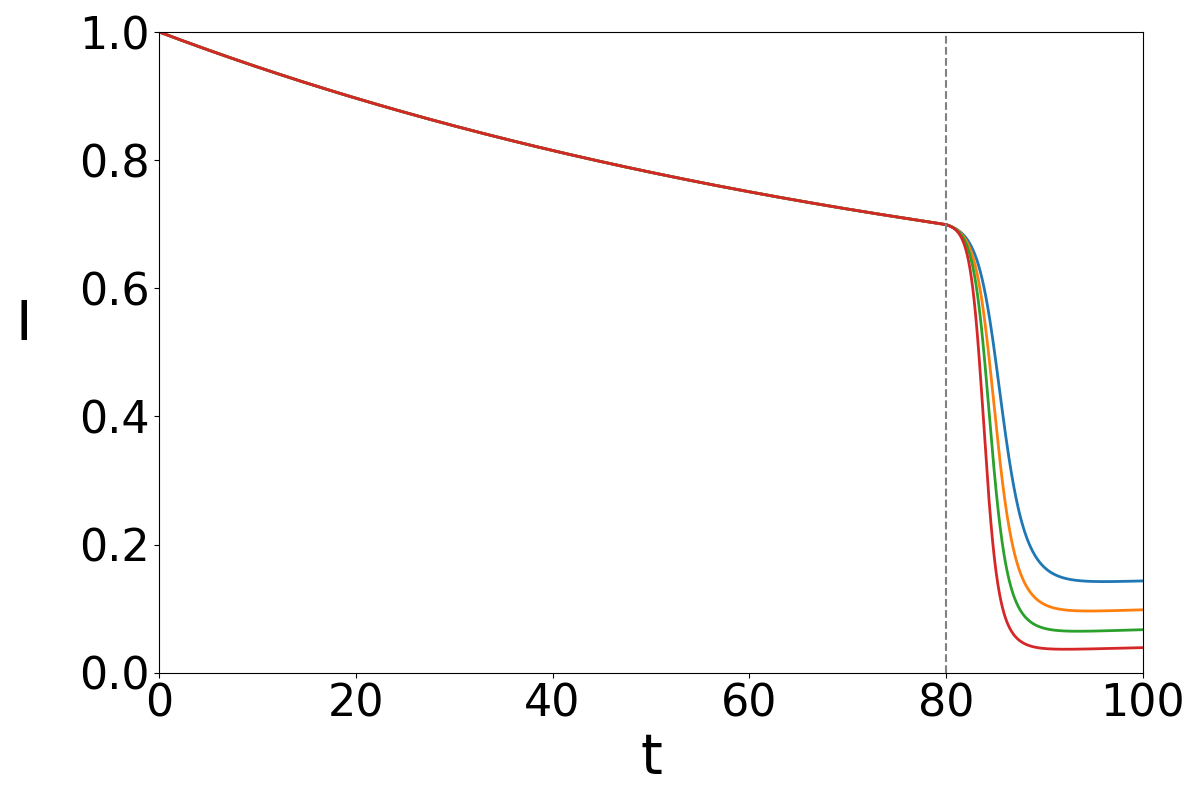}
        \caption{}
        \label{Fig:efficacy(a)}
    \end{subfigure}
    \hfill
    \begin{subfigure}[b]{0.24\textwidth}
        \centering
        \includegraphics[width=\linewidth]{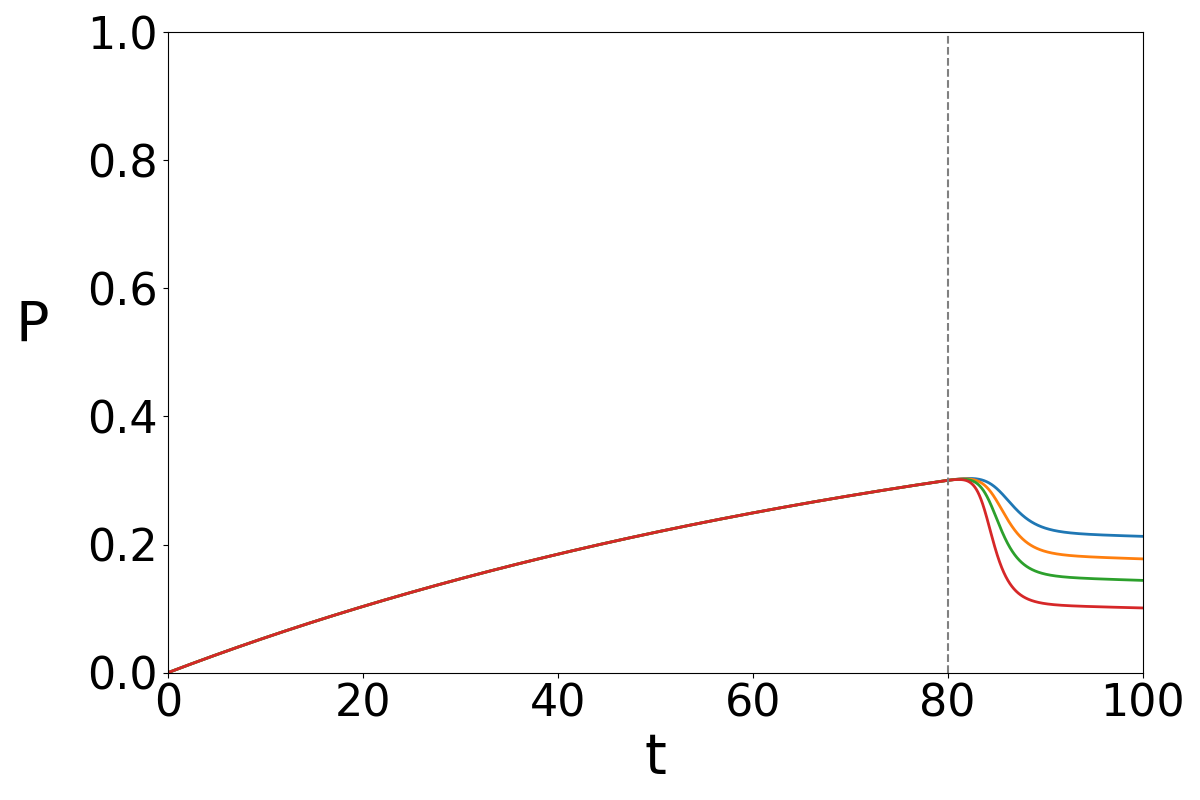}
        \caption{}
        \label{Fig:efficacy(b)}
    \end{subfigure}
    \hfill
    \begin{subfigure}[b]{0.25\textwidth}
        \centering
        \includegraphics[width=\linewidth]{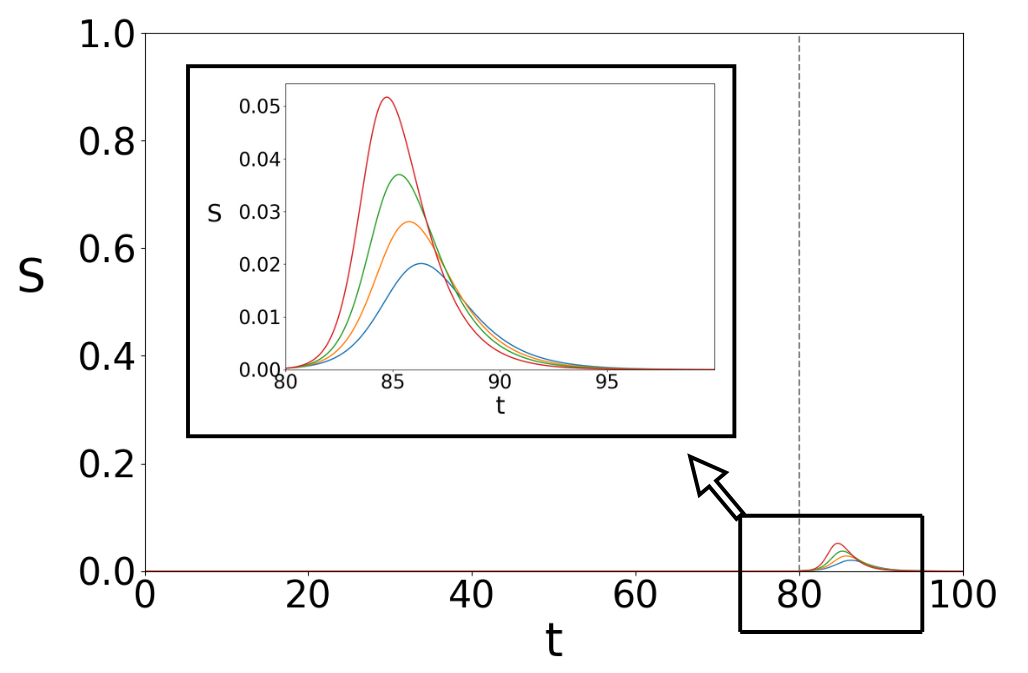}
        \caption{}
        \label{Fig:efficacy(c)}
    \end{subfigure}
    \hfill
    \begin{subfigure}[b]{0.24\textwidth}
        \centering
        \includegraphics[width=\linewidth]{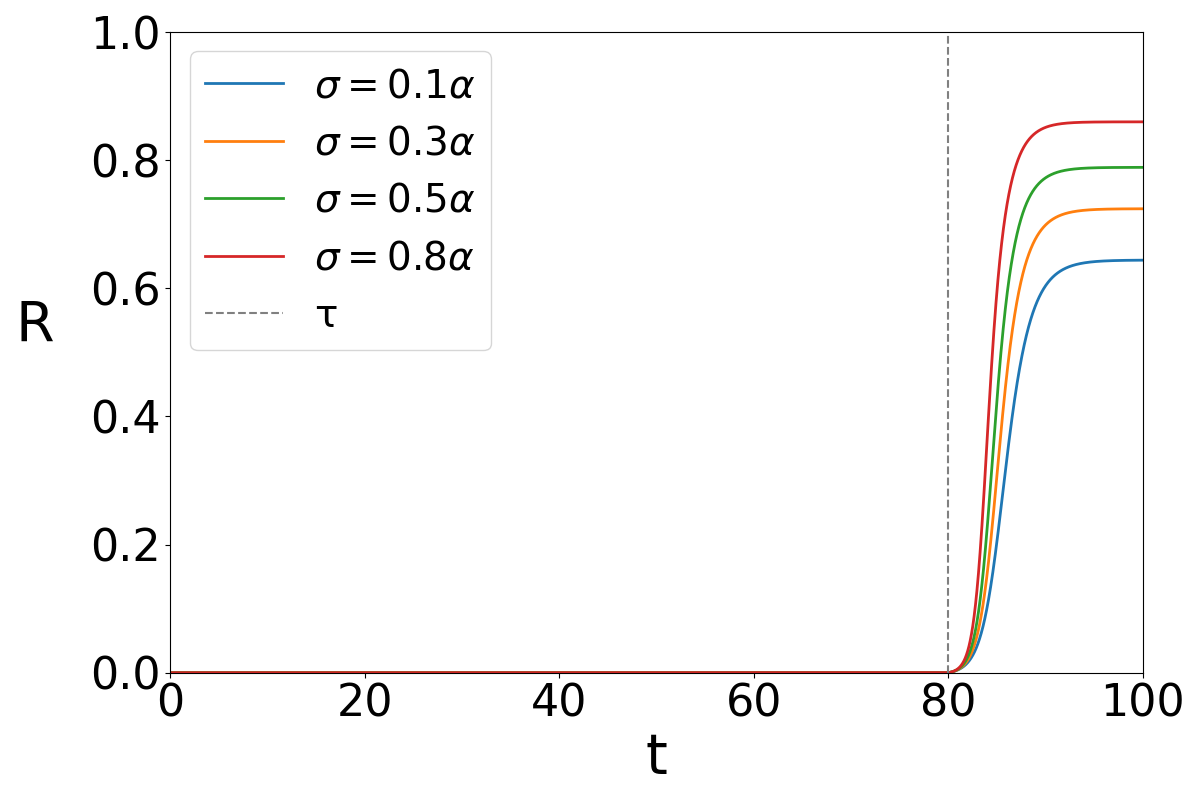}
        \caption{}
        \label{Fig:efficacy(d)}
    \end{subfigure}
    \caption{\textbf{Effect of \textit{prebunking} efficacy $(\sigma)$.} Influence of different $\sigma$ on the temporal evolution of \hspace{4in} (a) $I$, (b) $P$, (c) $S$ and (d) $R$.
    }
    \label{Fig:efficacy}
\end{figure*}
\subsection{Effect of \textit{prebunking} rate}\label{subsec41}
We study the effect of \textit{prebunking} rate $(\delta)$ on the dynamics of the IPSR model. The difference in $\delta$ values result to different populations in the Prebunked class at the start of misinformation spread, $t=\tau$. Keeping all parameters the same, Fig.\ref{Fig:rate} shows the dynamics of the different classes for various values of $\delta$.
The population of Ignorant class declines to reach a steady value after $t>\tau$. There is a minute difference in the Ignorant population irrespective of the change in $\delta$ (Fig.\ref{Fig:rate(a)}). For larger $\delta$, the Prebunked population at $t=\tau$ is larger which then declines to saturate to a larger value (Fig.\ref{Fig:rate(b)}). 
The $S$ class is $0$ for $t<\tau$. When $t>\tau$, $S$ increase rapidly to obtain a peak value and then it gradually declines to zero. $S$ contains only a small fraction of the total population and thus it is magnified for $t>\tau$ to clearly see the differences as the parameter is varied(Fig.\ref{Fig:rate(c)}). The peak number of spreaders is lower when $\delta$ is larger. Although we observe a high peak in the spreader population for low values of $\delta$, there is a larger rapid decline in the spreader population after the peak. Additionally, Fig. \ref{Fig:rate(d)} shows that larger values of $\delta$ leads to a significant reduction in the final population of stiflers. Overall, it indicates that larger \textit{prebunking} rate gives a larger fraction of the population in the prebunked class that can significantly limits the spread of misinformation, emphasizing the necessity of widespread \textit{prebunking} efforts.

\subsection{Effect of \textit{prebunking} efficacy}\label{subsec42}
We simulate how variation in \textit{prebunking} effectiveness $(\sigma)$ affects the dynamics of the IPSR model(Fig.\ref{Fig:efficacy}). This is achieved by using different values of $\sigma$, which represent the rate at which prebunked individuals convert to spreaders. The value of $\sigma$ is taken in fractions of $\alpha$ that represent different level of efficacies, with a fully susceptible population and no \textit{prebunking} effect when $\sigma=\alpha$.

\begin{figure}[h]
    \centering
    \includegraphics[width=0.9\linewidth]{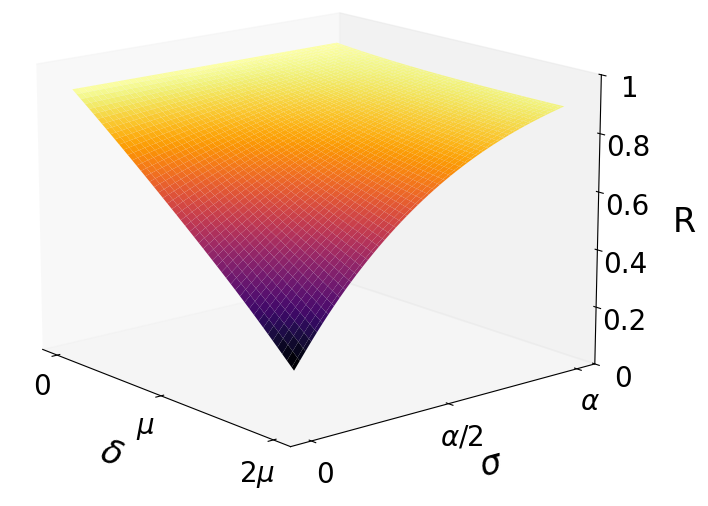}
    \caption{\textbf{The final scale of stifler as a function of $\delta$ and $\sigma$}. Here, $\delta \in [0,2\mu]$ and $\sigma \in [0, \alpha]$.}
    \label{Fig:final_scale}
\end{figure}
\begin{figure*}[htbp]
    \centering
    \begin{subfigure}[b]{0.32\textwidth}
        \centering
        \includegraphics[width=\linewidth]{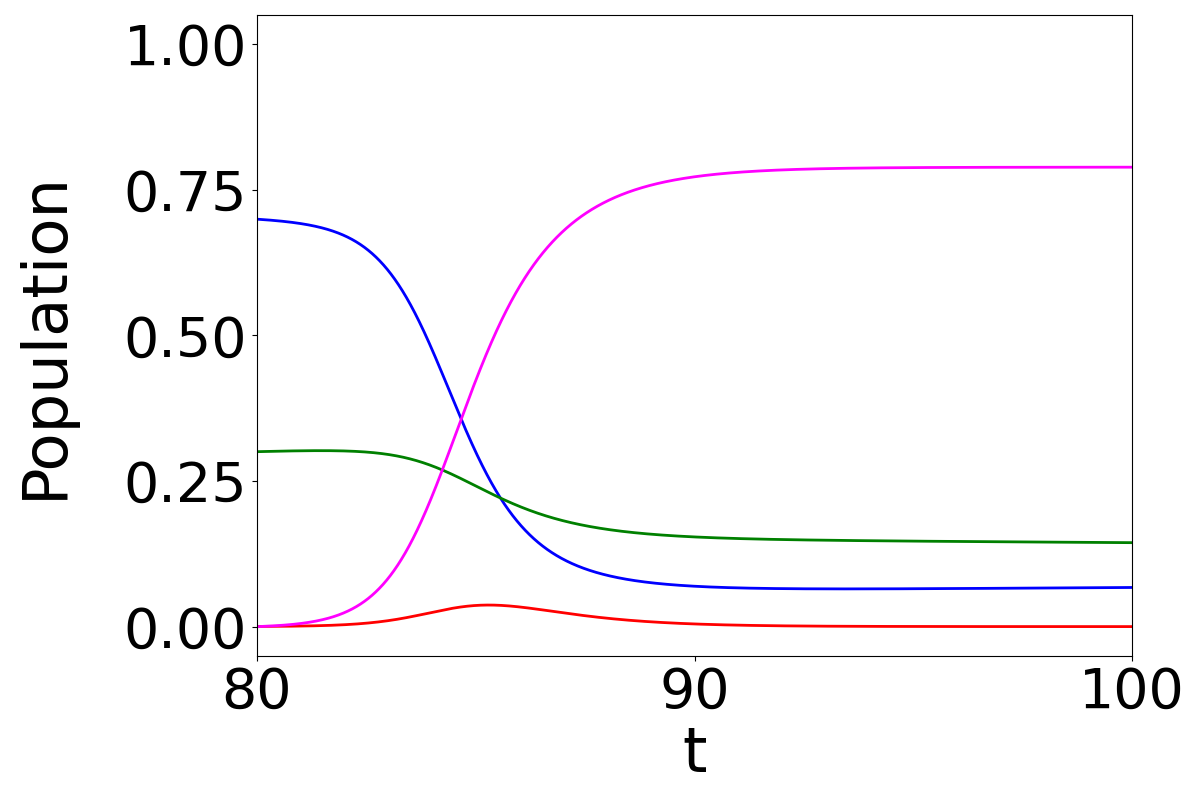}
        \caption{}
        \label{Fig:densities(a)}
    \end{subfigure}
    \hfill
    \begin{subfigure}[b]{0.32\textwidth}
        \centering
        \includegraphics[width=\linewidth]{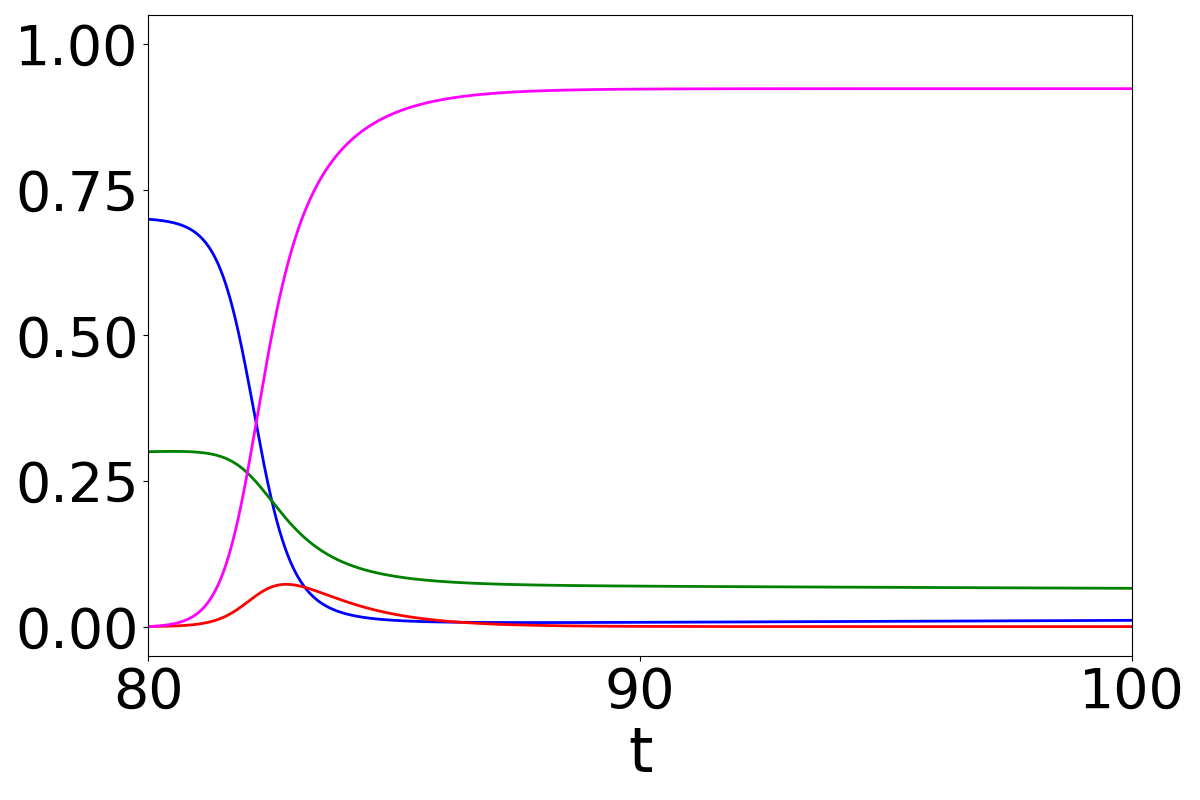}
        \caption{}
        \label{Fig:densities(b)}
    \end{subfigure}
    \hfill
    \begin{subfigure}[b]{0.32\textwidth}
        \centering
        \includegraphics[width=\linewidth]{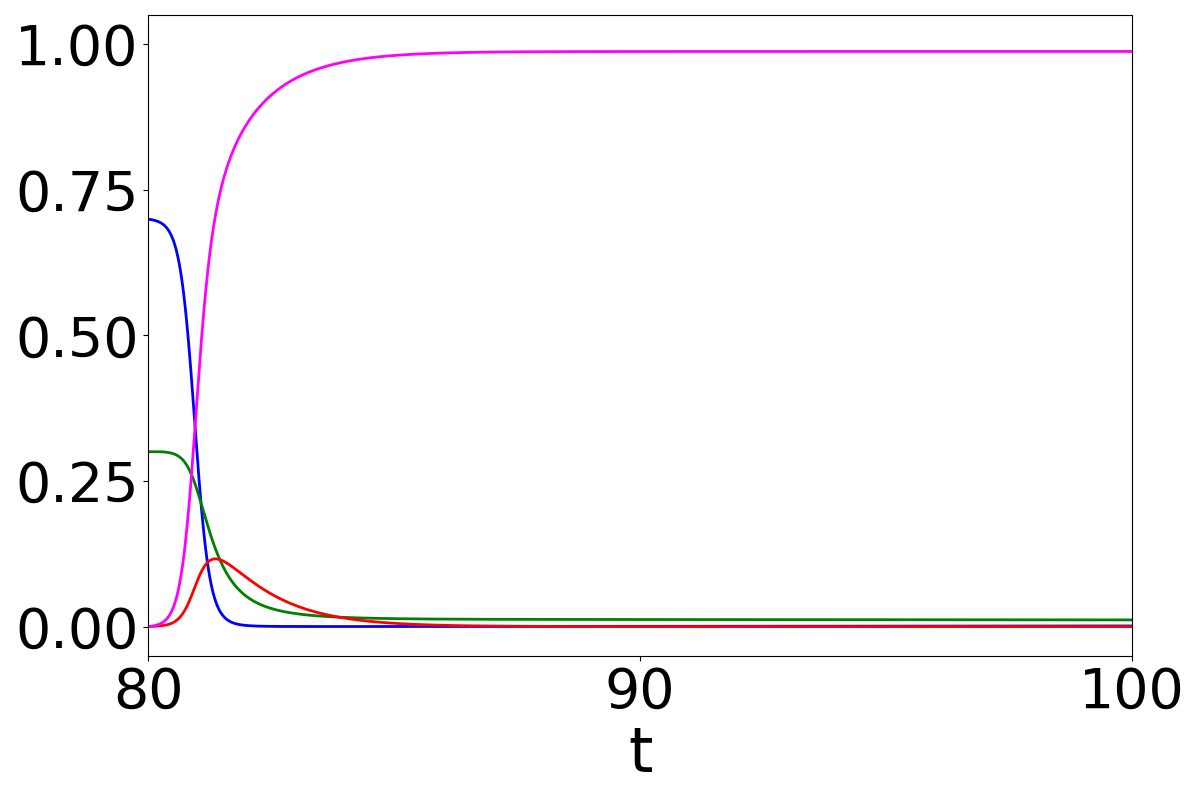}
        \caption{}
        \label{Fig:densities(c)}
    \end{subfigure}
    \caption{\textbf{Dynamics for different network densities} : Classes of IPSR model when $t>\tau$ for the values \hspace{5em} (a) $\langle k \rangle = 3$ \hspace{3em}   (b) $\langle k \rangle = 5$  \hspace{3em}  (c) $\langle k \rangle = 10$.
    }
    \label{Fig:densities}
\end{figure*}

In Fig.\ref{Fig:efficacy(a)}, the decay in the ignorant population when $t>\tau$ is rapid. As $\sigma \to \alpha$, the Ignorant settles at a lower population. The prebunked population (Fig. \ref{Fig:efficacy(b)}) also decreases from the initial prebunked value at $t=\tau$ and settles to a final value with the same trend as Fig.\ref{Fig:efficacy(a)}. In Fig.\ref{Fig:efficacy(c)}, we also find that low values of $\sigma$ results in a lower peak value of $S$ and a reduction in the final population of $R$ (Fig.\ref{Fig:efficacy(d)}). 
\textit{Prebunking}, when implemented, should be tailored with effective measures to overall decrease the scope of misinformation.

A 3D plot ( Fig.\ref{Fig:final_scale}) illustrates the steady state(final scale) of R for varying parameters of $\sigma$ and $\delta$. Larger \textit{prebunking} rate $\delta$ results to a larger prebunked population at $t=\tau$, combined with its effectiveness $\sigma$, can significantly reduce the final scale of stiflers. For values of $\sigma \to \alpha$ or $\delta \to 0$, the final stifler population increases, converging towards the value of the ISR model. This convergence is also reflected in the basic reproduction number(Eq.\ref{repro_formula}), when the system exhibits this approaching behaviour. 

\subsection{Influence of Network Density}
Real social networks, in general, are dense with a large average degree. To understand the influence of network density, we numerically simulate the time-evolution for the various classes of the IPSR model on 3 different values of $\langle k \rangle$ (Fig.\ref{Fig:densities}). The figures contain only the timeframe for $t> \tau$ to observe the misinformation spreading dynamics. The parameters remain consistent with the previous plots.

For higher values of $\langle k \rangle $, the system dynamics evolves more rapidly and converges to the equilibrium state much faster. The $I$ and $P$ classes diminish to a minimal value. The final scale of $R$ is much larger for denser networks and almost the whole population remains in this class. The peak value of Spreaders can also be seen to be higher for networks with higher $\langle k \rangle$. For the same initially prebunked population and \textit{prebunking} efficacy, denser networks are more resilient to \textit{prebunking} intervention and these networks facilitate faster and deeper percolation of misinformation within the network.

\subsection{Robustness of the model}
\begin{figure}[h]
    \centering
    \captionsetup{justification=justified, singlelinecheck=false}
    \includegraphics[width=0.9\linewidth]{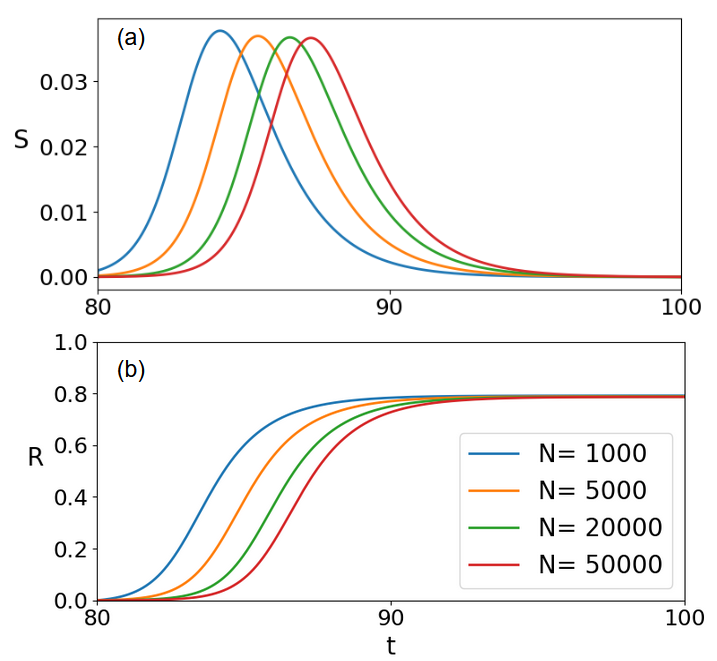}
    \caption{The dynamics of \textbf{(a) S} and \textbf{(b) R} of the IPSR model on different degree-homogeneous network sizes $N$. }
    \label{Fig:robust}
\end{figure}

We evaluate the robustness of the IPSR model by performing simulations on different degree-homogeneous network sizes with the same $\langle k \rangle$. Fig.\ref{Fig:robust} displays the Spreader and Stifler classes for varying network sizes. The prebunked population at $\tau = 80$ is the same as in earlier observations and it is maintained at $30\%$ of the total population. The other parameters are kept consistent with the earlier plots. We observe that the peak values of Spreaders for all network sizes remain approximately the same, although the temporal evolution is slower in larger networks as these networks are more sparse than smaller networks for the same $\langle k\rangle$. Similar distinctions in the rate of spreading is observed in the Stifler class for different $N$. However the final steady scale of $ R$ remains the same for all networks. Thus the model remains consistent for any network size.

\section{Dynamics on Heterogeneous Networks}\label{sec6}
In heterogeneous networks, nodes exhibit a broad range of degrees, with social networks often following small-world property and a power-law degree distribution. The mean-field approximation cannot accurately describe the dynamics of such structures. The IPSR model equations for such networks are redefined to include the network structure and node-level interactions, given in Appendix \ref{secA5}.

\begin{figure}
    \centering
    \captionsetup{justification=justified, singlelinecheck=off}
    \includegraphics[width=0.9\linewidth]{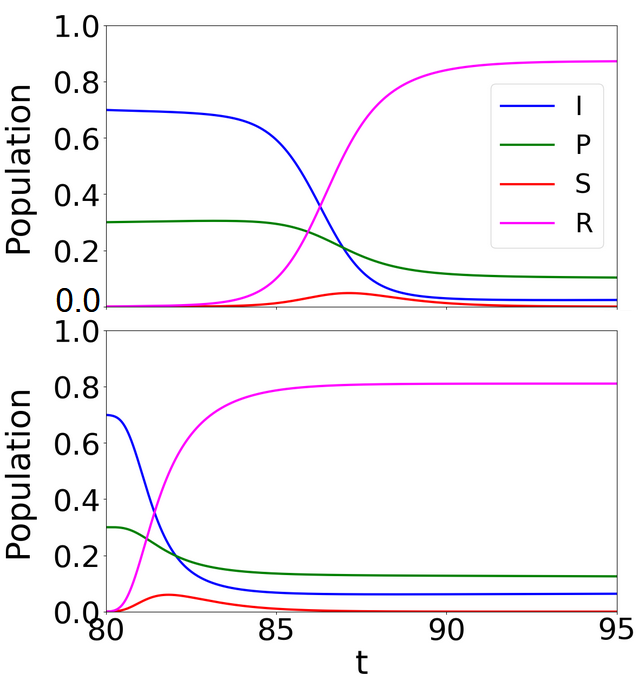}
    \caption{\textbf{IPSR Dynamics on Heterogeneous Networks.}
    (a) Watts-Strogatz network with $\langle k \rangle = 4$, $p = 0.1$;
    (b) Albert-Barabási network with $\langle m \rangle = 2$.
    Both networks have size $N = 3909$ and a single spreader at $\tau = 80$.}
    \label{Fig:networks}
\end{figure}

We numerically simulate the IPSR model on two types of synthetic networks: Watts–Strogatz (WS) and Barabasi–Albert (BA), each consisting of $N=3909$ nodes, corresponding to the number of users involved in the Charlie Hebdo event from the PHEME dataset. Both networks are generated using the NetworkX package in Python and are configured to have an average degree $\langle k \rangle \approx 4$. The WS network exhibits small-world properties such as short average path lengths and high clustering, while the BA network follows a power-law degree distribution, capturing the heterogeneity of node connectivity often seen in real-world social networks. Fig.\ref{Fig:networks} shows the simulated spreading dynamics with the same parameter values obtained from the empirical fitting. The plots display the dynamics for $t> \tau$ to focus on the active phase of misinformation spreading. In both networks, the fraction of Spreaders initially increases, reaches a peak, and then gradually declines. The temporal evolution of the Ignorant and Stifler populations follows trends similar to those observed in earlier results.
Notably, the BA network reaches its equilibrium state more rapidly than the WS network. The final fraction of Stifler's $R$ in the WS network exceeds slightly in comparison to the BA network. Despite this, the BA network shows a faster initial rise in $R$, driven by the presence of highly connected hub nodes. These central nodes facilitate rapid early dissemination of misinformation but also contribute to a swift transition to the Stifler state, thus curbing further spread. We observed similar conclusions on these networks while studying the dependence of parameters on the system, thus, we did not include those results in this paper.

\section{Conclusion and Discussion}\label{sec7}
This work employs a compartmental model to investigate the impact of \textit{prebunking} intervention on the spread of misinformation. Initially, we introduce the ISR model for understanding the dynamics of information spreading. We estimate the parameters of the ISR model using a real world rumor dataset from Twitter. Then, we formulate the IPSR model to account for both the \textit{prebunking} efforts and the forgetting mechanism of \textit{prebunking} over time. In the IPSR, the population is categorized into four distinct compartments: ignorant, prebunked, spreader, and stifler, based on their exposure to \textit{prebunking} and misinformation. 
We derive an expression for the basic reproduction number and identify the conditions for the propagation of information, as well as the conditions that prevent the emergence of a significant number of new spreaders. We analytically determine the steady states for the IPSR model and the stability condition of these steady states.

Further, we conduct numerical simulations to examine the dynamics of the IPSR model. We apply a series of sensitivity analyses, yielding various conditions for \textit{prebunking} to reduce the scope of misinformation in a social system. A substantial reduction in the number of spreaders and stiflers is observed for certain parametric conditions. Thus, to reduce the effects of misinformation outbreaks, an effective strategy is to educate people about the different tactics of misinformation and develop cognitive immunization through \textit{prebunking}.

To our knowledge, our work is the first to propose a macroscopic model for misinformation intervention that incorporates \textit{prebunking}. The mean-field equations capture the dynamics of the population within a system that has a degree-homogeneous network. Social networks typically exhibit a power-law degree distribution characterized by hubs and nodes with a wide range of degrees, resulting in a heterogeneous network structure. Thus we also simulated on WS and BA networks to verify the consistency of our results. Real-world data with \textit{prebunking} efforts is currently unavailable because researchers and practitioners have only recently begun to implement and study \textit{prebunking} techniques. Hence, we use the PHEME dataset from Twitter to study spreading dynamics and employ intervention using \textit{prebunking}.

Although achieving $100\%$ \textit{prebunking} of the population may not be possible, focusing on reaching the largest possible fraction with effective efforts to counter misinformation is crucial.
Effective \textit{prebunking} can enhance people's cognitive resilience to misinformation, but individuals often forget the awareness, making them vulnerable to misinformation over time. This increases the risk of spreading misinformation to a larger population. Therefore, it is crucial to conduct timely \textit{prebunking} efforts before significant events, such as elections or pandemics, to mitigate the impact of misinformation; otherwise, the forgetting mechanism can fully hamper the benefits of \textit{prebunking} and allow misinformation to spread unchecked.

In future research, we will examine the impact of hubs and communities on the overall effectiveness of \textit{prebunking} in intervening misinformation the dissemination. Additionally, studies using microscopic models such as the Independent Cascade and Linear Threshold Models that account for \textit{prebunking} would be valuable. The IPSR model allow for a deeper understanding of information dynamics and the development of more effective \textit{prebunking} strategies.

\bmhead{Acknowledgements}
CM acknowledges support from the Anusandhan National Research Foundation (ANRF) India (Grants Numbers SRG/2023/001846 and EEQ/2023/001080).

\bmhead{Data Availability Statement} The PHEME dataset from Twitter is available at \url{https://doi.org/10.6084/m9.figshare.2068650.v2}. The data that support the findings in this study are available from the author upon request.

\appendix

\section{Proof of theorem 1}\label{secA1}
\begin{proof}
    From the Eq. \ref{2a}, 
    \begin{equation*}
        \frac{dI}{dt} = -\delta I -(\alpha + \gamma) \langle k \rangle I S + \mu P
    \end{equation*}
    which holds true 
    \begin{align*}
        \frac{dI}{dt} &\geq -\delta I -(\alpha + \gamma) \langle k \rangle I S \\
        \frac{dI}{I} &\geq -(\delta + (\alpha + \gamma) \langle k \rangle S) dt\\
        \int \frac{dI}{I} &\geq  -\int(\delta + (\alpha + \gamma) \langle k \rangle S) dt\\
        I(t) &\geq I(0) e^{-\delta t - (\alpha+\gamma) \langle k \rangle \int S dt } \geq 0.
    \end{align*}
    Similarly, %Taking the second sub-equation from
    Eq. \ref{2b} holds true,
    \begin{align*}
        %\frac{dP}{dt} &= \delta I -(\sigma + \rho) \langle k \rangle P S - \mu P \\
        \frac{dP}{dt} &\geq -(\sigma + \rho) \langle k \rangle P S - \mu P\\
        \frac{dP}{P} &\geq -((\sigma + \rho) \langle k \rangle S + \mu) dt\\
        \int \frac{dP}{P} &\geq  -\int((\sigma + \rho) \langle k \rangle S + \mu) dt\\
        P(t) &\geq P(0) e^{-\mu t - (\sigma+\rho) \langle k \rangle \int S dt } \geq 0
    \end{align*}
    Similarly, taking the Eq. \ref{2c},
    \begin{align*}
        \frac{dS}{dt} &= \alpha \langle k \rangle I S + \sigma \langle k \rangle P S - \beta S \\
        \frac{dS}{dt} &\geq -\beta S\\
        \frac{dS}{S} &\geq -\beta dt\\
        S(t) &\geq S(0) e^{-\beta t} \geq 0
    \end{align*}
    The terms in the Eq. \ref{2d} are all positive and confirm that,
    \begin{equation*}
        \frac{dR}{dt} \geq 0.
    \end{equation*}
    Since all components of the model yield non-negative solutions, we can conclude that the system solution remains positive for all $t>0$.
\end{proof}

\section{Proof of theorem 2}\label{secA2}
\begin{proof}
    To determine the steady state of the system, we set all four sub-equations from Eq. \ref{md:model} to zero.
    \begin{equation*}
        \mathlarger{\frac{dI}{dt}}=0, \quad \mathlarger{\frac{dP}{dt}}=0, \quad  \mathlarger{\frac{dP}{dt}}=0,\quad \mathlarger{\frac{dR}{dt}}=0
    \end{equation*}
        
    This yields $ S=0$ and $\delta I = \mu P $, which is the proposed straightforward solution for the system to be in an equilibrium state.  
\end{proof}
The case with $I,P,S=0$ is trivial, which is not considered, as it results in $R=1$. 
This situation is flawed because, in real-world scenarios, misinformation completely dies out before reaching the entire population $(R\neq 1)$.

\section{Proof of theorem 3}\label{secA3}
\begin{proof}
    At the equilibrium condition $S=0$, the Jacobian matrix of the system takes the following form:
    \begin{equation}
    J = \begin{pmatrix}
    -\delta & \mu & -(\alpha + \gamma)\langle k \rangle I & 0  \\
    \delta & -\mu & -(\sigma+\rho)\langle k \rangle P & 0  \\
    0 & 0 & \alpha\langle k \rangle I +\sigma\langle k \rangle P -\beta & 0  \\

    0 & 0 & \rho\langle k \rangle P+\gamma\langle k \rangle I+\beta & 0
    \end{pmatrix}
    \label{C1}
    \end{equation}

    Two of its eigenvalues as zero ($\lambda_1, \lambda_2=0$), and the other two eigenvalues are 
    \begin{equation*}
        \lambda_3 = -(\delta+\mu) \hspace{4em} \lambda_4 = \alpha\langle k \rangle I+\sigma\langle k \rangle P-\beta
    \end{equation*}
    By applying the second condition from \textbf{theorem 2}, the last eigenvalue $\lambda_4$ can be expressed as follows: 
    \begin{equation*}
        \lambda_4 = \left({\frac{\alpha\mu+\sigma\delta}{\delta}}\right)\langle k \rangle P-\beta
    \end{equation*}
    \vspace{0.3em}
    Given that the parameters are assumed to be non-zero and positive, the $\lambda_3$ will be negative. For the proposed condition (Eq. \ref{C2}), $\lambda_4$ will be negative. %indicates stable equilibrium when the proposed condition is satisfied.
\end{proof}
The largest eigenvalue of the Jacobian matrix (Eq. \ref{C1}) is zero, so we further use the Lyapunov function to analyze the stability of the steady state of the system.
\section{Proof of theorem 4}\label{secA4}
\begin{proof}
    We establish the global asymptotic stability of the IPSR model by introducing a Lyapunov function and prove that it exhibits monotonicity along the system’s trajectories. 
     Consider the Lyapunov function 
    \begin{equation*}
        L  = \frac{1}{2} (I + P)^2
    \end{equation*}
    This function is positive definite for positive non-zero parameters.
    The time derivative of the Lyapunov function yields
    \begin{align*}
    L' &= (I+P)\Bigg(\frac{dI}{dt} + \frac{dP}{dt} \Bigg)\\
    &= (I+P) \bigg[ -\delta I - (\alpha + \gamma)\langle k\rangle I S +\mu P \\
    & \quad \quad \quad \quad \quad +  \delta I - (\sigma + \rho)\langle k\rangle P S -\mu P  \bigg] \\
    & = (I+P) \bigg[- (\alpha + \gamma)\langle k\rangle I S  - (\sigma + \rho)\langle k\rangle P S   \bigg]
    \end{align*}
    which is strictly negative for all positive non-zero parameters.\\
    Also, we can see that $L' = 0$ under the equilibrium condition $S=0$.
\end{proof}

\section{IPSR model equations for heterogeneous networks}\label{secA5}
Heterogeneous networks are characterized by the presence of highly connected hubs and nodes with varying degrees. If $A$ is the adjacency matrix of a network and assuming equal susceptibility of individuals to misinformation and \textit{prebunking}, the equations that drive the dynamics of the IPSR model are 
\begin{subequations}
    \begin{align*}
    \frac{dI_i}{dt} &= -\delta I_i -(\alpha + \gamma) I_i \sum_{j} A_{ij}S_j + \mu P_i\\
    \frac{dP_i}{dt} &= \delta I_i -(\sigma + \rho) P_i \sum_{j} A_{ij} S_j - \mu P_i \\
    \frac{dS_i}{dt} &= \alpha I_i\sum_{j} A_{ij} S_j + \sigma P_i\sum_{j} A_{ij} S_j - \beta S_i\\
    \frac{dR_i}{dt} &= \rho P_i\sum_{j} A_{ij} S_j + \gamma I_i \sum_{j} A_{ij} S_j +  \beta S_i
    \end{align*}
\end{subequations} 
where $I_i$, $P_i$, $S_i$, $R_i$ are the probabilities that node $i$ is in that class. The transition rates correspond to those defined earlier in the case of homogeneous networks.
\nocite{*}
\bibliography{references}

\end{document}